# Valley Polarization and Inversion in Strained Graphene via Pseudo-Landau Levels, Valley Splitting of Real Landau Levels and Confined States

Si-Yu Li[1,§], Ying Su[2,§], Ya-Ning Ren[1], and Lin He[1,3,*]

[1] Center for Advanced Quantum Studies, Department of Physics, Beijing Normal University, Beijing, 100875, People's Republic of China

[2] Theoretical Division, T-4 and CNLS, Los Alamos National Laboratory, Los Alamos, New Mexico 87545, USA

[3] State Key Laboratory of Functional Materials for Informatics, Shanghai Institute of Microsystem and Information Technology, Chinese Academy of Sciences, 865 Changning Road, Shanghai 200050, People's Republic of China

[§]These authors contributed equally to this work.
*Correspondence and requests for materials should be addressed to L.H. (e-mail: helin@bnu.edu.cn).

**It is quite easy to control spin polarization and spin direction of a system via magnetic fields. However, there is no such a direct and efficient way to manipulate valley pseudospin degree of freedom. Here, we demonstrate experimentally that it is possible to realize valley polarization and valley inversion in graphene by using both strain-induced pseudomagnetic fields and real magnetic fields. The pseudomagnetic fields, which are quite different from real magnetic fields, pointing in opposite directions at the two distinct valleys of graphene. Therefore, coexistence of the pseudomagnetic fields and the real magnetic fields leads to imbalanced effective magnetic fields at two distinct valleys of graphene. This allows us to control the valley in graphene as convenient as the electron spin. In this work, we report consistent observation of valley polarization and inversion in strained graphene via pseudo-Landau levels, splitting of real Landau levels and**

**valley splitting of confined states using scanning tunneling spectroscopy. Our results highlight a pathway to valleytronics in strained graphene-based platforms.**

Besides the spin degree of freedom, quasiparticles in graphene offer another degree of freedom, the valley quantum number, which can be described as a two-level system in analogy to the spin[1-12]. Previous scanning tunneling microscopy (STM) studies demonstrated that it is possible to lift the valley degeneracy in graphene by introducing symmetry-breaking potential from substrates[13,14] or electron-electron interactions[15,16]. Then, one can measure the valley polarization through valley-split Landau levels (LLs) in the presence of real magnetic fields[13-16]. Very recently, it was demonstrated that we can detect the valley splitting in graphene at nanoscale and single-electron level by confining the valley-split LLs in STM-tip-induced quantum dots (QDs)[17-19]. Both valley polarization and inversion are observed in the confined states of the graphene QDs because of strains and broken-symmetry potential arising from the supporting boron nitride[18]. Theoretically, there is an alternative route, *i.e.*, by introducing both pseudomagnetic field (PMF) and real magnetic field, to realize valley polarization in graphene in a more controllable way. The PMF, arising from modulation of electron hopping due to lattice deformation, can effectively create Landau quantization around low-energy Dirac cones of graphene[20-24]. Such a pseudo-Landau quantization has been explicitly demonstrated by STM measurements in which the pseudo-LLs are clearly observed[25-35]. However, the PMF, unlike the real magnetic field, does not violate the time-reversal symmetry of graphene and has opposite signs in the $K_+$ and $K_-$ valleys[20-

[24]. Therefore, coexistence of the PMFs and the real magnetic fields leads to imbalanced effective magnetic fields at two distinct valleys[36-44], which allows us to realize valley polarized Landau quantization in graphene[29].

In this Letter, we report STM measurements on a folded area of graphene and demonstrate that there is PMF gradient along the fold, including an inversion point where the pseudo-LLs disappear, *i.e.*, the PMF becomes zero. In the presence of real magnetic fields, we observe valley-polarized LLs along the fold and the observed valley splitting inverted at the inversion point. By confining the LLs into STM-tip-induced QDs, we further detect the valley polarization and inversion of the confined states along the fold at single-electron level. Our experiment indicate that coupled PMF and real magnetic field can be used to switch valley polarization and inversion in graphene, which is extremely important in the valleytronics.

In our experiment, we carry out measurements on graphene multilayers grown on Rh foils. By applying strong perpendicular magnetic fields, we can observe well-defined Landau quantization of massless Dirac fermions (Fig. 1), indicating that the topmost graphene sheet is electronically decoupled from the supporting substrate. Previously, we have demonstrated that mismatch of thermal expansion coefficients between the graphene and the metal substrates can introduce ripples in graphene[26,27,29,45-47]. In our experiment, nanoscale ripples can be easily observed in the topmost graphene sheet (see Fig. S1 for STM images[48]). Therefore, such a system provides an ideal platform to study novel electronic properties induced by the PMF in graphene monolayer.

Figure 1(a) shows a representative STM image of a graphene nanoripple (see Fig. S2

for more atomic-resolved STM images[48]). In strained graphene, it is quite difficult to generate uniform PMFs[20-24]. Our experimental results, as shown in Fig. 1(b) and 1(c), indicate that the PMFs along the ripple depend sensitively on the measured positions, even when the structure of the ripple only changes slightly (see Fig. S3 for detailed analysis of the strain along the ripple[48]). Figure 1(c) shows three representative scanning tunneling spectroscopy (STS) spectra acquired in zero magnetic field on different positions along the ripple. The tunneling peaks in the spectra are attributed to the strain-induced pseudo-LLs, which can be fitted well by the Landau quantization of massless Dirac fermions[13-16,62-65] (see Fig. S4[48]). According to the fitting, we can obtain the value of the PMFs in each position of the ripple. Figure 1(b) shows distribution of PMFs measured along the arrow in Fig. 1(a). Obviously, the PMFs are quite non-uniform and there is a very narrow transition region (less than 2 nm, within two dashed lines in Fig. 1(a)) where the PMF is almost zero.

To further understand origin of the spatial distribution of the PMFs, we analyze lattice deformation of the strained graphene and carry out theoretical calculation. Through fast Fourier transform (FFT) analysis of the atomic-resolved STM images, we obtain the lattice deformation at different positions along the ripple (Fig. S3[48]), which indicates that there are compressive strains along the zigzag direction both below and above the narrow transition region. Therefore, we can infer an in-plane deformation associated with an out-of-plane deformation of the graphene ripple. The strained graphene can be described by the tight-binding Hamiltonian[22]:

$$H = -t \sum_{\langle i,j \rangle} \left(1 - \frac{\beta}{a} d_{ij}\right) e^{i\frac{e}{\hbar} \int_i^j \mathbf{A} \cdot d\mathbf{r}} c_i^\dagger c_j$$

where $t = 2.7$ eV is the hopping energy and $a = 0.142$ nm is the bond length between nearest neighboring (NN) carbon atoms in pristine graphene. Due to the deformation, the bond length between NN sites $i$ and $j$ is changed by $d_{ij}$, and $\beta = -\partial \ln t/\partial \ln a \approx 2$. Here $\boldsymbol{A}$ is the vector potential induced by the external magnetic field. The exact strain field due to the deformation is hardly determined from the STM image with limited resolution. Instead, we propose a simple ansatz to approximate the strain field, which captures the leading order effects of the deformation (see Fig. S7-10 for details[48]). Our theoretical calculation by taking into account the spatial variation of strain shows that the direction of PMF is reversed along the ridge of the ripple (Fig. 2(a))[66]. Thus a narrow boundary with zero PMF exists naturally at the transition point. Therefore, our experimental results, complemented by theoretical calculations, demonstrate that the PMF has opposite signs in the two adjacent regions in the rippled structure shown in Fig. 1(a).

The PMF exactly has the same value but has opposite signs in the two valleys (Fig. 2(a)). By applying a real magnetic field $B$, the total effective magnetic field in the two valleys becomes different. In graphene monolayer, the energies of $n \neq 0$ LLs depend on the magnetic fields and only the $n = 0$ LL is independent of the magnetic fields. Therefore, we can realize valley polarization for the $n \neq 0$ LLs by using both the PMFs and real magnetic fields. Figure 1(d) shows three representative spectra recorded in different positions of the ripple in $B = 11$ T. We can observe well-defined Landau quantization of graphene monolayer. There is a pronounced asymmetry between the intensities of the LLs with positive and negative orbital index. This may partly arise from the decrease of quasiparticle lifetimes with the energy difference from the Fermi energy[15] and partly arise from large electron-hole asymmetry in strained graphene[29,46]. Besides the pronounced asymmetry, the other obvious feature of the spectra is the clear splitting of $n$ = -1, -2 and -3 LLs along with undetectable splitting of the zero LL recorded at the positions with finite PMFs, as shown in Fig. 1(d) (see Figs. S5 and S6 for more spectra and discussion[48]). At the boundary with zero PMF, the splitting of all the LLs vanishes. These results demonstrate explicitly that the splitting of the LLs is induced by the combined effect of the PMFs and the real magnetic field.

Our calculation indicates that the coexistence of PMFs and real magnetic fields leads to the emergence of valley polarization for the $n \neq 0$ LLs, as shown in Fig. 2(b) and Fig. 2(c). In our experiment, the valley splittings of the $n$ = -2, -3 LLs are not as pronounced as that of the $n$ = -1 LL (Fig. 1(d)). Such a phenomenon is also well reproduced in our numerical result (Fig. 2(b) and 2(c)). Theoretically, the description of modulated electron hopping in strained graphene as an effective PMF is exactly valid only at the charge neutrality point. Apart from the charge neutrality point, the high pseudo-LLs are not well defined and the low-energy effective PMFs description fails. Therefore, we can obtain better valley splitting with decreasing LL index for the $n \neq 0$ LLs. Because of the reverse of the PMF around the boundary in the studied ripple, the valley polarization changes signs from positive in the top region to negative in the bottom region (Fig. 1(d), Fig. 2(b) and Fig. 2(c)). Here positive (negative) valley polarization is defined as the energy of the LLs in the $K_+$ valley is larger (smaller) than that in the $K_-$ valley. At the boundary, the valley polarization is zero, which is independent of the real magnetic fields. The above result demonstrates explicitly that the coupled PMFs and real magnetic field can be used to switch valley polarization and valley inversion in graphene.

To further measure the spatially varying valley splitting in strained graphene, we use a recently developed method based on edge-free graphene quantum dots (GQDs), as schematically shown in Fig. 3(a). The edge-free GQD is generated by combining the electric field of the STM tip with a perpendicular magnetic field[17-19]. The probing STM tip, acting as a moveable top gate, bends the LLs in the region beneath the tip into the gaps between the LLs, which leads to edge-free confinement in graphene and generates confined orbital states in the GQDs (Fig. 3(a)). The orbital states of the GQDs can reflect the degeneracies of electrons in the studied graphene regions[17-19]. In pristine graphene, there are four-fold degeneracies (spin and valley) for electronic states. Therefore, every single orbital state of the GQD could be occupied by four electrons (See Fig. S11). Because of the small capacitance $C$ of the GQD, a single excess electron on the GQD needs to overcome the electrostatic energy $E_c = e^2/C$. As a consequence, we can observe a series of quadruplets of charging peaks in STS spectrum of pristine graphene (See Fig. S12). When the valley degeneracy is lifted, every single quadruplet of the confined orbital states will be divided into two doublets, as schematically shown in Fig. 3(b).

The GQDs is movable with the STM tip, consequently, we can detect the valley splitting $E_V$ of graphene at any position where we chose. Figure 3(c) shows representative STS spectra recorded on the orange dot in Fig. 1(a) in different magnetic fields. Besides the well-defined LLs at low bias, we also observed charging peaks at high bias, indicating the emergence of the GQD beneath the STM tip. Obviously, the four charging peaks of a bound state divided into two doublets. In our experiment, the first four charging peaks arise from the confinement of the $n = -1$ LL. We define the energy spacing as $\Delta E_{12}$, $\Delta E_{23}$ and $\Delta E_{34}$, with $\Delta E_{12} = E_c + E_Z$, $\Delta E_{23} = E_c + E_V - E_Z$, and $\Delta E_{34} = E_c + E_Z$ respectively ($E_Z = g\mu_B B$ with the effective g-factor $g \approx 2$ is the Zeeman splitting). Here the energy spacing of the confined states $\Delta E$ can be directly deduced from the voltage difference $\Delta V_{tip}$ acquired from the charging peaks in the STS spectra by using $\Delta E = \eta e \Delta V_{tip}$ with $\eta$ as the tip lever arm (See Figs. S13 and S14 for more analysis of the $\eta$)[17-19]. The valley splitting $\Delta E_V$ of a selected position can be directly acquired according to $\Delta E_V = \Delta E_{23} - \Delta E_{12} + 2E_Z$. In Fig. 3(d), we summarize the valley splitting $\Delta E_V$ measured in the orange dot of Fig. 1(a) as a function of $B$. In the meanwhile, we also plot the valley splitting of the same position deduced from the splitting of the $n = -1$ LL in Fig. 3(d). Obviously, the valley splitting measured by the two methods agrees well with each other, further demonstrating that it is facile to detect the valley splitting at nanoscale by using the GQD. Figure 3(e) shows the four charging peaks measured at different positions along the arrow in Fig. 1(a). It indicates that the valley splitting depends on the measured positions and reverses the direction around the boundary where the PMF is zero.

To clearly show the spatially varying valley splitting, we measured 70×70 spectra in the studied area shown in Fig. 4(a) at a fixed $B$. Then we can obtain the values of $\Delta E_{12}$ ($\Delta E_{34}$), $\Delta E_{23}$ and $|\Delta E_V|$ as a function of positions in the studied area. Figures 4(b)-

4(d) show representative results obtained at $B = 11$ T and we obtain the valley splitting in the whole studied area (Fig. 4(d)). Similar results are obtained in other different $B$ (see Fig. S15). Obviously, the valley splitting is observed over a wide range of the rippled region, implying that the strain field is rather homogeneous over regions more extended than the magnetic lengths of both the pseudomagnetic fields and real magnetic fields. Therefore, we can observe well defined Landau quantization, as shown in Fig. 1. Figure 4(e) summarizes the valley splitting as a function of positions measured along the arrow shown in Fig. 1(a) in different $B$. Obviously, there is a narrow boundary with zero valley splitting that separates two adjacent areas with opposite valley splitting. The maximum valley splitting changes from about 20 meV in the top region to -10 meV in the bottom region in the studied nanoripple. Here we should point out that the non-strained transition region separating two adjacent regions with the compressive strain is critical for the observation of the valley inversion in our experiment. In strained graphene with non-uniform compressive strain but without the non-strained region, we still can observe spatially varying valley splitting due to the non-uniform of the PMFs. However, the valley splitting cannot decrease to zero and reverse its direction (see Fig. S16-S18 for more experimental data and analysis).

The above experiment indicates that the local strain of graphene plays a vital role in determining the valley polarization and inversion. Therefore, it is very important to control the local strain of graphene. Very recently, our experiment demonstrates that it is possible to control the local strain to generate programmable graphene nanobubbles by using functional atomic force microscopy (AFM)[35]. The size and shape of the graphene nanobubbles can be tuned by the stimulus bias of AFM tip. In this work, although it is still difficult to precisely control the local strain of graphene, we show the ability to change the local strain and, consequently, the corresponding valley polarization of graphene by using STM tip (Fig. S19-21). In the near future, it is possible to control strain patterns in graphene by using AFM or STM tip.

In summary, we report consistent observation of valley polarization and inversion in strained graphene via pseudo-LLs, splitting of real LLs and valley splitting of confined states in tip induced QDs. Similar to the Zeeman splitting of electron spins, the valley degeneracy can be lifted by the imbalanced effective magnetic fields at the two distinct valleys of graphene. The resulted valley-polarized LL splitting up to several tens meV provides an ideal two-level subsystem formed purely by valley polarized quantum states. Our results provide a new avenue to manipulate the valley, which may have potential applications in valleytronics.

**Acknowledgments**
This work was supported by the National Natural Science Foundation of China (Grant Nos. 11974050, 11674029). L.H. also acknowledges support from the National Program for Support of Top-notch Young Professionals, support from "the Fundamental Research Funds for the Central Universities", and support from "Chang Jiang Scholars Program". Y.S. was supported by the U.S. Department of Energy through the Los Alamos National Laboratory LDRD program, and was supported by the Center for Non-linear Studies at LANL.

In our experiment, the observed pseudo-LLs and valley-split LLs are not limited in one ripple. As shown in Fig. 4(c) and 4(d), the Landau level splitting is observed over a wide range of the rippled region that is larger than the magnetic lengths. Therefore, we can observe well defined quantizations.

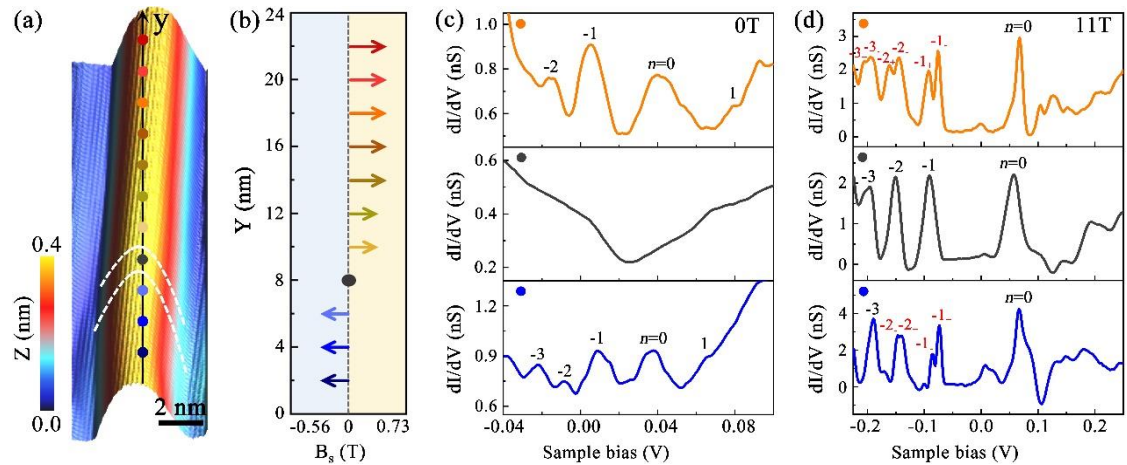

**Figure 1.** (a) A STM image ($V_{sample}$ = 600 mV and $I$ = 0.3 nA) of a nanoscale ripple. The black arrow and the dots in different colors mark the positions where we take the STS spectra. In the region between the two dashed curves, the measured PMF is zero and we cannot detect valley-polarized LLs. (b) Measured PMFs along the arrow in panel a. (c) Three tunnelling spectra measured at different positions in panel a. The indexes for the pseudo-Landau levels are marked. (d) Three tunnelling spectra measured at different positions in panel a in $B$ = 11 T. The indexes for the LLs are marked. The subscribe +/- represents LLs of the $K_+$/$K_-$ valley.

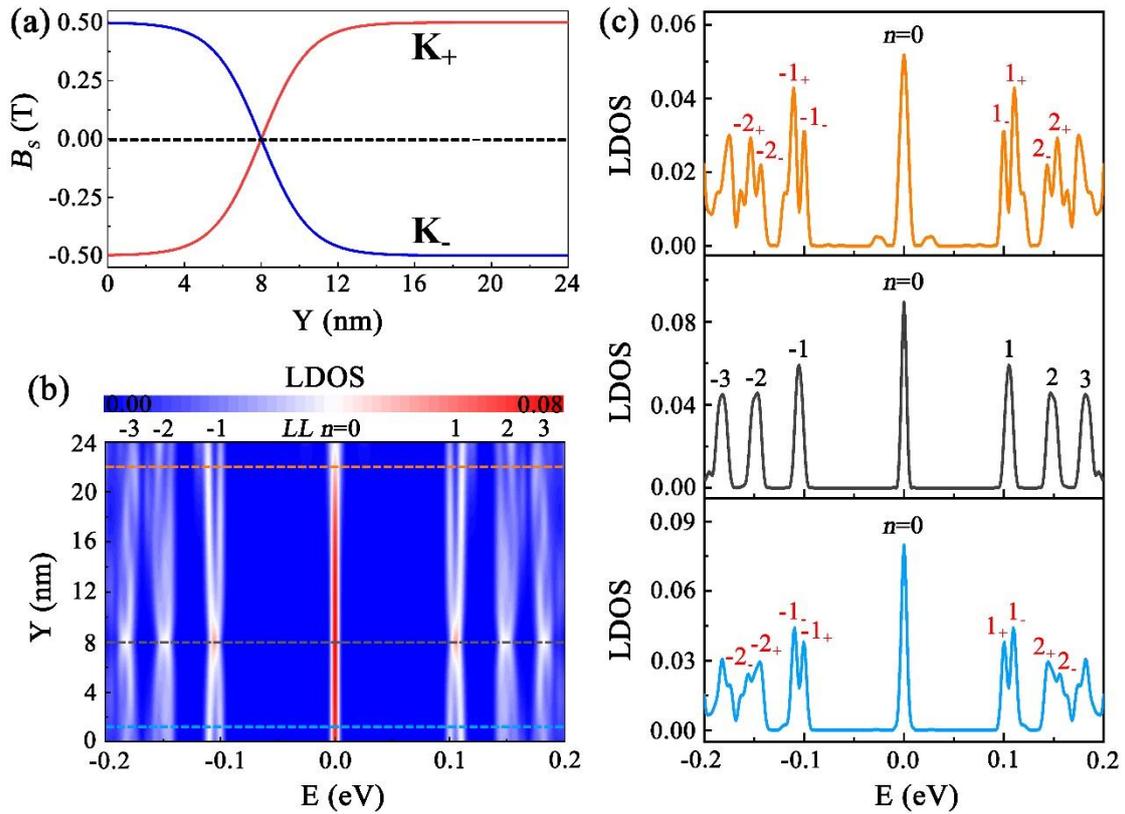

**Figure 2.** (a) Calculated PMFs along the ridge of the ripple at the $K_+$ and $K_-$ valleys. PMF is zero at the crossing point. (b) LDOS along the ridge of the ripple in $B$ = 11 T. Splittings of the ±1 and ±2 LLs are clearly shown around the two ends of the ripple. (c) The LDOS at three different positions, corresponding to the three dashed lines (from up to down) in panel b.

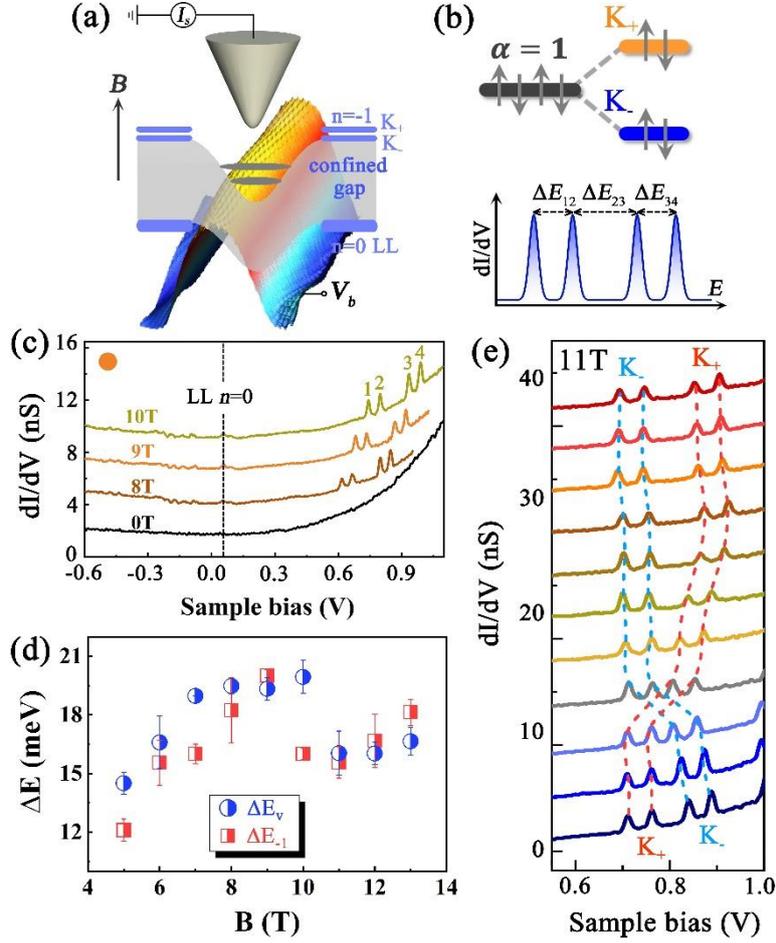

**Figure 3.** (a) Schematic of the tip-induced edge-free GQDs with the valley-polarized LL in external magnetic field. The STM tip leads to the band bending beneath the tip to realize the edge-free GQDs. (b) Top: Schematic valley-split confined state in the GQD. Bottom: Schematic for a sequence of doublet charging peaks in the *dI/dV* spectrum. (c) The STS spectra taken at the orange dot in Fig. 1(a) in different magnetic fields. The spectra are offset in Y axis for clarity. The numbers 1-4 sign the sequence of doublet charging peaks. (d) The valley splitting in different magnetic fields, yielding from the sequence of charging peaks ($\Delta E_V$) and the splitting of the LL$_{-1}$ ($\Delta E_{-1}$), respectively. (e) 11 STS spectra taken along the arrow of Fig. 1(a) in $B = 11$ T. Valley polarization and inversion are clearly observed.

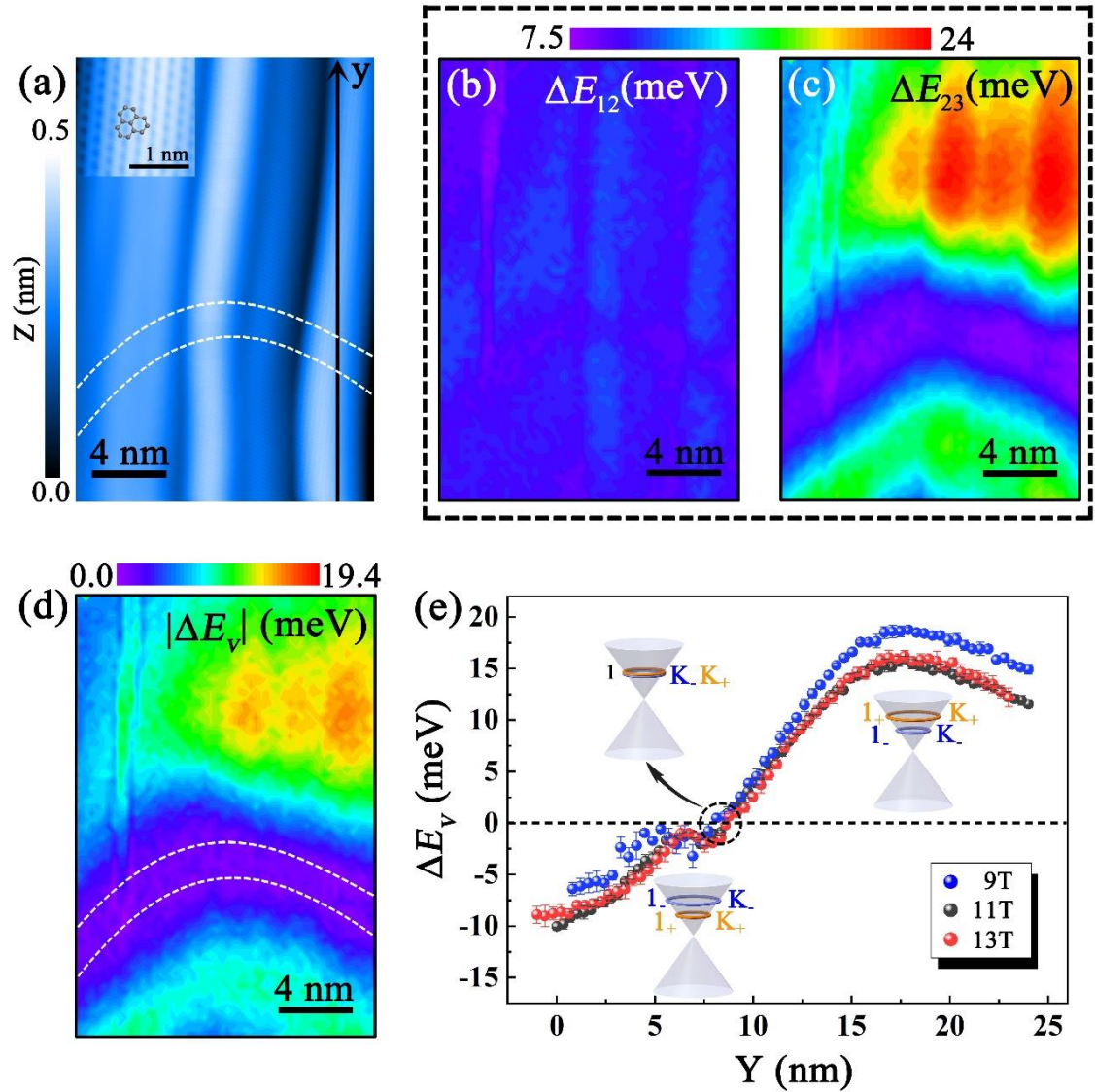

**Figure 4.** (a) A STM image ($V_{sample}$ = 400 mV and $I$ = 0.3 nA) of nanoscale graphene ripples. In the region between the two dashed curves, the measured PMF is zero and we cannot detect valley-polarized LLs. (b-d) Energy maps for the values of $\Delta E_{12}$, $\Delta E_{23}$ and $|\Delta E_V|$ as a function of positions in the studied area in 11 T. (e) The valley splitting as a function of positions measured along the arrow shown in Fig. 1(a) in different magnetic fields. The insets schematically show the valley splitting and valley inversion.